# Embedded Systems Education in the 2020s: Challenges, Reflections, and Future Directions

Sudeep Pasricha
Department of Electrical and Computer Engineering, Colorado State University
Fort Collins, CO, USA
sudeep@colostate.edu

## ABSTRACT

Embedded computing systems are pervasive in our everyday lives, imparting digital intelligence to a variety of electronic platforms used in our vehicles, smart appliances, wearables, mobile devices, and computers. The need to train the next generation of embedded systems designers and engineers with relevant skills across hardware, software, and their co-design remains pressing today. This paper describes the evolution of embedded systems education over the past two decades and challenges facing the designers and instructors of embedded systems curricula in the 2020s. Reflections from over a decade of teaching the design of embedded computing systems are presented, with insights on strategies that show promise to address these challenges. Lastly, some important future directions in embedded systems education are highlighted.

## 1. INTRODUCTION

The first modern electronic embedded systems can be traced back to the mid-1960s, when the advent of integrated circuits led to the design of the real-time Apollo Guidance Computer, developed at MIT for the Apollo Space Program [1]. Soon after, in 1969, Japanese company Busicom asked Intel to design a set of custom integrated circuits for use in a line of business calculators. The resulting Intel 4004 chip, introduced in 1971, was the world's first single-chip microprocessor, designed to read and execute a set of instructions (software) stored in an external memory chip [2]. Intel hoped that the software would give each calculator its unique set of features and that this design style would drive demand for its core business in memory chips. The 4004 was an overnight success, and its use increased steadily over the next decade. Early embedded systems applications included business calculators, computerized traffic lights, unmanned space probes, and aircraft flight control systems. Today, embedded systems have become ubiquitous across the medical, consumer, robotics, automotive, energy, manufacturing, and networking application domains. This rapid proliferation of embedded systems, with a market size that is expected to reach $1.4 trillion by 2026 [3], motivates the need for a workforce that is adequately trained to meet future industry needs.

An embedded system is a combination of computer hardware and software, and potentially additional mechanical or electronic parts, designed to perform one or more dedicated functions, either independently or as part of a larger system. In recent years, the phrase 'embedded systems' has been superseded in popularity by 'Internet of Things' (IoT). However, IoT devices are actually a subset of embedded systems that possess connectivity to the Internet. Another related phrase that is also frequently used is 'cyber-physical systems' (CPS), which refers to the study and design, in a broader context, of an embedded system (cyber system) together with the environment (physical system) it operates within. Interested readers are referred to [4] for a discussion on the biases and emphasis associated with these phrases. Regardless of the specific terminology used to refer to them, 'embedded systems' represents an interdisciplinary field that combines many areas in computer science (CS) and electrical engineering (EE). As these systems become more complex, distributed, and networked today, the challenges of educating future embedded systems designers that are proficient across multiple disciplines takes on great importance.

It is also crucial that the embedded systems educational process be aligned with the competences and abilities required by the embedded systems industry. A survey of embedded system employers [5] emphasized 'interdisciplinarity' as a mandatory requirement in the education of future embedded systems designers. It was mentioned that these designers must be competent with hardware and software components, and also hardware-software codesign with a horizontal set of disciplines that relate to present and future application areas. In particular, the ability to abstract from a problem to a model and then moving through the whole design chain down to implementation was seen as characterizing the 'ideal' embedded system designer. The surveyed employers also emphasized soft skills such as teamwork, management, and communication. A large survey of embedded systems developers conducted in 2019 [6] also confirms the need for such a broad skillset for future embedded system designers.

## 2. BACKGROUND

One of the earliest studies on embedded systems education was published in 2000 [7] and asserted that future embedded systems designers need to be trained to have knowledge of the entire design process so that they can make global rather than local decisions. The authors go on to say that "We believe that next-generation courses in embedded computing should move away from the discussion of components and toward the discussion of analysis and design of systems". Such an interdisciplinary and system-centric approach has guided embedded systems curricula in universities for the past two decades. Several excellent embedded systems textbooks that have been published by Marwadel [8], Vahid and Givargis [9], Lee and Seshia [10], and Wolf [11] have followed this system-centric approach. These books cover hardware and software components, and their co-design, often with real embedded systems case studies to illustrate the importance of different components.

Many studies have discussed experiences with designing embedded systems courses and the use of specific components and tools in labs to drive embedded systems education. Experiences with undergraduate embedded systems courses at Princeton University and the Danish Technical University were presented in [7]. The emphasis in these courses from the late 1990s and early 2000s was on C and assembly based microcontroller programming and foundations of concurrency, CPU hardware, I/O, and system design. An important milestone was the inaugural Workshop on Embedded Systems Education (WESE) in 2005 that has since been an important forum for discussions related to embedded systems education. The many excellent talks at the workshop have covered three major themes: 1) experiences with embedded systems courses, 2) design of embedded systems curricula at universities, and 3) case studies of using specific hardware and/or software components, development boards, and exploration tools in embedded systems courses.

In [12], guiding principles were presented for the embedded systems teaching and research agenda at UC Berkeley. The authors discuss how embedded systems concepts are used to bring together the physical and computational world and how students are educated in critical reasoning about modeling and abstraction. Experiences with an embedded systems course at KTH were presented in [13]. This study from the mid-2000s advocated for an 'everything of something' approach, with deep dives into a few chosen topics related to microcontroller programming, sensors/actuators, and I/O. There have been other overviews of similar embedded systems courses at Nagoya University in Japan [14], Hong Kong University of Science and Technology [15], and the University of Dortmund in Germany [16]. In [17], efforts made by the Korean government are presented with an aim to meet the increasing industrial demand for quality IT experts in the computer-software field, including embedded systems. A spiral model for curriculum development is described that includes product requirement, design, implementation and realization phases.

Many studies since the mid-2000s have also emphasized labs in embedded systems courses that help to explain specific embedded system design concepts, e.g., coverage of real-time principles with the help of a railroad control system project that utilized real model trains [18], real-time control via a project based on implementing a servo controller for a robotic arm [19], and hardware/software co-design via a project involving partitioning and mapping of a JPEG decoder onto an FPGA board [20]. In [21], experiences with embedded systems coursework at Columbia were presented, with a focus on using an FPGA board as a unifying component for labs, and teaching VHDL and C programming to enable the hands-on exercises in the labs. This FPGA-centric approach has been widely adopted in many embedded systems courses, e.g., as discussed in the 2013 study from National University of Singapore [22].

Since the early 2010s, with the growing importance of CPS, there has been a shift towards integrating content related to the interaction between cyber and physical components, as part of new CPS courses that also cover embedded systems, e.g., [23]. This shift can also be observed in the renaming of the WESE workshop to 'Workshop on Embedded and Cyber-Physical Systems Education' in 2012. Embedded systems courses in the past decade have also integrated more contemporary technologies, components, and case studies to stay relevant to the needs of embedded systems industry. For instance, many embedded systems courses have adopted open-source hardware boards, e.g., Raspberry Pi, BeagleBoard, Arduino, etc, to teach embedded systems programming and co-design. These boards have been used in sequences of embedded system-related courses, and also in undergraduate capstone projects, e.g., [24]. Embedded systems courses at UC Irvine [25] and others have adopted Android OS based smartphones and development boards to teach embedded systems programming and co-design. Other embedded systems courses have included hands-on projects with contemporary drones and readily-available robotics platforms, e.g., using LEGO Mindstorms to teach concurrent, real-time dependent, and networked embedded system software design [26]. Lastly, an interesting development is the computer engineering curriculum developed by a joint ACM/IEEE task force in 2016 [30] which allocates up to 40 core hours to the embedded systems area, highlighting its importance in contemporary computing education.

## 3. CHALLENGES

From the discussion in the previous section, it is clear that embedded systems education has had a vibrant history since the early 2000s. So, as of 2022, what are the challenges facing embedded systems education? Based on analysis in literature [5] as well as our own experience, several challenges remain:

**Complex system synthesis:** Embedded systems engineers must be familiar with general engineering methods which involve the design and realization of systems that adhere to specified functional and technical properties in a methodical manner. To design and synthesize complex systems, engineers have to work in diverse teams and decompose the entire system into appropriate subsystems such that the properties of the integrated system can be synthesized [29]. Structuring and decomposing a system into subsystems and dealing with the additional complexities of real-world implementations are very challenging for students, as these learning experiences are underrepresented in CS and EE curricula.

**Large content footprint:** Embedded systems have evolved in recent years to include increasingly autonomous decision-making, support ubiquitous networked operations (e.g., for IoT systems), and interact more comprehensively with the physical world (e.g., in CPS platforms). Thus, embedded system design is now much more complex than it was a few decades ago. The design challenges today also span many more scientific areas, and therefore embedded systems courses are faced with the problem of needing to cover an increasing number of topics within a limited amount of time. Past approaches in embedded system curricula that advocate covering 'everything of something' [13] (focusing on deep dives into a small number of topics) can lead to deficiencies in skillsets for future embedded systems designers trained through such curricula.

**Instruction expertise:** Covering the variety of topics in the scope of embedded system design requires expertise across hardware engineering, software engineering, hardware/software co-design, controls, optimization, networking, real-time systems, testing, security, and reliability. Very few instructors have deep knowledge across such a vast spectrum of topics. Software stacks and hardware technologies in embedded systems also change very rapidly, requiring frequent updates to labs and lecture material. Therefore, embedded system instruction must be simultaneously durable and practical [31]. If too much emphasis is placed on how to achieve design goals with today's technology, then students receive technical training with only short-term value. If there is too much emphasis on foundational theory, students gain no intuition about the physical realities of designing complex embedded systems.

**Student background:** Student "readiness" for the breadth and depth required across embedded systems topics is certainly a major challenge. Students from CS and EE backgrounds each have a specific set of core competencies and skills, often with a limited overlap [32], yet both backgrounds are essential to comprehend embedded system-specific topics. A considerable number of EE students are not very comfortable with software programming, whereas many CS students have little motivation or knowledge to design hardware. Soft skills are also not given as much emphasis in most curricula, which together with deficiencies in socio-emotional competence, can lead to some students failing to successfully navigate the heavy demands of embedded systems coursework.

**Student motivation:** Improving student motivation to achieve ambitious learning goals is an important goal for all curricula, and thus also for embedded systems courses. The self-determination theory in the educational context [33] serves as a theoretical framework to enhance motivation among university students. The self-determination theory [34] maintains that a person has three innate needs: 1) The need for autonomy, i.e., the need to feel that the individual's behavior was not forced upon them, but emanates from their needs; 2) The need for competence, i.e., the need to feel that the individual is capable, and can meet challenging objectives; and 3) The need for relatedness, i.e., the need to love and be loved, and to be a part of a group. When a student's needs are satisfied, this will bring them to a high level of motivation, while the prevention of this satisfaction will compromise it. Further, the theory postulates that the origin of motivation is a spectrum spanning extrinsic factors (e.g., due to fear of punishment or failure, hopes of receiving remuneration) and intrinsic factors (e.g., emanating from interest or pleasure). The theory states that as much as motivation emanates from more intrinsic factors, the higher its quality will be. Embedded systems curricula must be designed to maximize such motivation.

**Pandemic and distance education:** Due to unexpected disruptions such as those witnessed with the recent COVID-19 pandemic, there have been pedagogical shifts that impact embedded systems education. There has been a growing reliance on virtual lectures and online instruction due to isolation requirements and international travel restrictions. This makes it challenging to conduct embedded systems labs and assignments that require physical equipment to be accessed by students. Online/distance instruction is also on the rise due to returning learners from industry who want to update their skills but also work full time and cannot attend lectures in-person. These trends raise new questions about the most effective methods to engage and motivate such students in embedded systems courses.

The next section describes reflections from over a decade of teaching the design of embedded systems at Colorado State University (CSU) [35]. The application-specific pedagogical approach taken for the coursework has shown to be beneficial in addressing some of the important challenges highlighted above.

## 4. REFLECTIONS ON MODERN EMBEDDED SYSTEMS COURSE DESIGN

I have designed and taught an embedded systems course [35] at CSU since 2009. This is an introductory/intermediate level embedded systems course targeting junior and senior undergraduate students, and first year graduate students in the CS and ECE departments. Students enrolling in the course are required to have taken an introductory course on microcontrollers. At CSU, such courses exist in both the CS and ECE departments at the freshman and sophomore levels, e.g., 'Introduction to Microprocessors', which covers C and assembly programming to control peripherals (e.g., displays, sensors, servo motors) with ARM processor-based boards such as the Raspberry Pi.

The pedagogical objectives of this course are: 1) provide end-to-end skills needed to understand and design embedded systems today and in the forseeable future, and 2) align course content with the theory of self-determination to improve student motivation, participation, and learning. Much like the field of embedded systems, the course content has evolved over time. But a primary thread that has remained consistent is an emphasis on student-driven application-based embedded systems projects. This approach is a departure from the board-centric approach taken in many embedded systems courses where labs and projects center around a specific hardware board or platform. Students in the course are encouraged to come up with applications that are of interest to them and prototype an embedded system for those applications by the end of the semester-long course.

Several examples of relevant applications are provided to students as a starting point, including embedded systems use-cases for 'home security', 'indoor navigation with IoT devices', 'autonomous drones', 'smart musical instruments', 'agricultural control', etc. Recommendations are provided for the problem scope in these applications, along with suggestions for open-source hardware boards, software components, tools, and peripherals. The students are required to select an application area of focus and submit a proposal outlining their plan for the embedded systems prototyping within the first three weeks of the course.

The course content in lectures, homework/labs, and reading assignments is structured to support the students' efforts towards prototyping an embedded system for their selected applications. Fig. 1 outlines the main modules in the course taught over a span of 16 weeks. Table 1 describes the content covered across the modules in more detail. The final week (Week 17) is reserved for interactive application-driven project presentations and prototype demos from student teams, and the final exam.

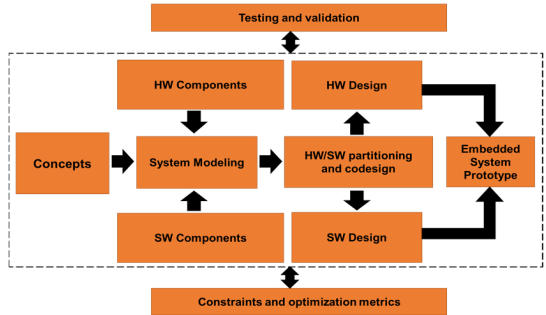

**Figure 1: Outline of the embedded systems course modules**

**Table 1: Course content across modules over the entire semester**

| Week | Content |
|---|---|
| 1 | Overview of fundamental concepts related to embedded systems, IoT, and CPS platforms; real-time systems; cross-layer design; application domains, time-to-market economics |
| 2-3 | Models of computation and specification languages for simulating and analyzing embedded systems; deep dives into SystemC for HW/SW modeling and co-design; optional tutorials on VHDL/Verilog via pre-recorded modules. |
| 4-6 | Embedded system software design; deep dives into embedded system-specific software engineering principles (e.g., waterfall, agile models and tools for project management and collaborative development), algorithmic complexity, real-time scheduling and mapping techniques for computation (e.g., tasks) and communication (e.g., messages on CAN buses), real-time operating systems (RTOS), and device drivers. |
| 7-10 | Embedded system hardware design; deep dives into embedded hardware processors including DSPs, GPUs, TPUs, and FPGAs, as well as memory, network, and peripherals; hardware component selection, interfacing, integration, and optimization; scratchpad caches, networking standards (e.g., CAN/FlexRay/Ethernet, IEEE 802 RF standards), digital signal processing, domain-specific accelerators. |
| 11 | HW/SW partitioning algorithms and heuristics; multi-objective optimization and trade-offs across relevant metrics, e.g., energy, power, performance, cost, and area. |
| 12 | Embedded system software design and optimization; best practices for writing safety-critical embedded system software (including coverage of MISRA guidelines); code compression techniques; compilation optimizations; testing/validation. |
| 13 | Embedded system hardware design optimization; logic synthesis; tools for architectural evaluation, optimizations for fault-tolerance and predictable computing; testing/validation. |
| 14 | Sensors, actuators, A2D/D2A conversion, embedded control; challenges with quantization, aliasing, noise, and calibration. |
| 15 | Secure embedded systems; fundamentals of cryptography, embedded system attack case studies, techniques to secure HW/SW subsystems with appropriate overheads. |
| 16 | Case studies of designing applications with embedded systems in medical, automotive, and consumer domains. |

Over the 13 years that my course has been taught at CSU, there have been a number of insights gained, based on instructor experience and student assessment feedback:

**Application-driven approach:** Based on early offerings of the course that either required working with a specific board, or only involved survey reports on embedded systems-related themes as end-of-semester deliverables, students seem to prefer and enjoy the application-driven approach taken in the course. Fig. 2 shows a collage of some of the creative projects completed by students in the course. Allowing students the flexibility to focus on a single application domain problem of their interest, and providing the supportive content to allow them to select, interface, and optimize embedded system components toward a prototype seems to have had a very positive impact on student participation and engagement throughout the duration of the course. Table 2 shows student survey scores for selected years, with the increase in student satisfaction corresponding to the adoption of the application-driven approach by 2013. Student readiness and background limitations to undertake prototyping for complex embedded applications were addressed by encouraging EE and CS students to pair together into groups (cross-listing the course across these departments helped diversify enrollment), to complement each other's skills. Requiring students to work in teams, present their work, and demonstrate prototypes also provided students experience with some of the soft skills needed in the embedded system industry, and which the students may not get elsewhere (e.g., CS students at CSU do not get to sufficiently practice such soft skills due to lack of capstone projects, unlike EE students). Thus, the application-driven approach has shown promise to address the student-related challenges discussed in Section 3.

**Mixed-mode instruction:** The course was one of the first courses in the ECE department at CSU to be offered in a mixed-mode format, with lectures simultaneously recorded for later viewing and also streamed online in real-time, since 2009. But engaging distance students is not easy due to many factors, e.g., due to some of them preferring to watch lectures asynchronously, working time-consuming full-time jobs, being unable to network with their peers in person, etc. Working on embedded system boards and in teams is also difficult for these students. Fortunately, the application-driven approach seems to work well for distance students as it does not require shipping specific boards to them or requiring them to procure specific platforms that may not be easily available (especially as some distance students who took the course lived outside the USA). Best practices over a decade of online instruction in the course helped improve distance student engagement, including regularly monitored course-specific online discussion forums, online Q&A sessions with the teaching assistant and instructors, icebreaker events to engage in-class and online students, etc. These practices have helped with the pandemic/distance education-related challenges discussed in Section 3.

**Course content, assignments, labs:** The course content has changed considerably over time. In its current form, it aims to balance theory and practical knowledge needed to prototype embedded systems, as well as embedded system-related standards and emerging directions that are desirable for preparing the next generation of embedded system designers. Students seem to generally appreciate this balance (see Table 2). Homework/labs in the course have been designed with the same goals of balancing theory (e.g., with questions on scheduling theory) and practice (e.g., optimizing real-world embedded software), and involve working with open-source and free software tools (e.g., GEM5 architectural simulator, Android OS). Reading assignments provide deeper insights into each module (Fig. 1 and Table 1). Regular low-stakes quizzes/surveys allow for timely assessment, constructive feedback to students, and corrective adjustments to course content over the semester. These approaches for content design, assignments and labs, and assessments have helped address some of the content- and instruction-related challenges discussed in Section 3.

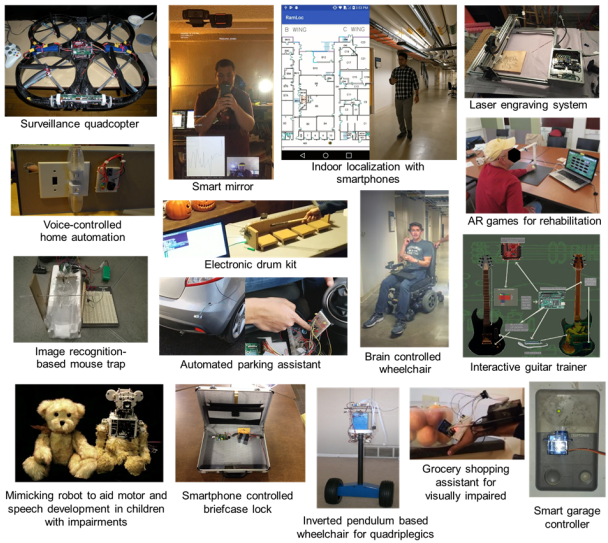

Figure 2: Examples of various student-defined application-driven embedded systems projects in the course since 2009.

Table 2: Course evaluation (numbers are average scores; 1: poor, 2: below average, 3: average, 4: very good, 5: excellent)

|  | 2009 | 2013 | 2017 | 2021 |
|---|---|---|---|---|
| Enrollment | 30 | 55 | 65 | 38 |
| How well did class lectures/sessions increase your understanding of the subject? | 3.43 | 4.35 | 4.71 | 4.65 |
| How well did assignments/projects increase your understanding of the subject? | 3.82 | 4.39 | 4.38 | 4.60 |
| How do you rate this course? | 3.61 | 4.57 | 4.67 | 4.54 |

**Student motivation:** The improvement in intrinsic motivation carries great importance, as embedded systems curricula requires the development of higher order thinking skills, and intrinsic motivation has a central role in this level of learning [33]. On a deeper level, the high level of motivation expressed by students at the end of the course may be explained by the self-determination theory, according to which the satisfaction of the individual's needs increases motivation. The need for autonomy was satisfied in the course by the independence granted to the students during the application-driven project. The need for competence was realized by the focused guidance from the instructor and complementary labs/assignments/lectures throughout the semester. Finally, the need for relatedness was satisfied by students working in teams with their peers and the personal attitude of the course instructor.

**Reuse vs. development:** It is important to find the right balance between students developing their own code and reusing existing code. The embedded systems developers survey from 2019 [6] indicated that as much as 88% of software code and 77% hardware IP in embedded systems projects is reused. Thus, learning to exploit reuse opportunities is an important skill. However, significant reuse also diminishes experiences with developing and validating new components from scratch. Students in the course were encouraged to reuse open-source hardware and software components whenever possible. However, it was also conveyed to students that they need to create new modules in their project that were entirely their own contribution. Students were required to clearly document reuse and new development during project proposals and mid-semester checkpoint reports, with feedback provided by the instructor on the appropriate balance between reuse and new development.

**Student success lessons:** A majority of students demonstrated the ability to work in a very structured manner, by identifying relevant applications and sub-problems, and then decomposing the overall engineering objectives accordingly. However, I note that the kind of autonomy and style of working involved with the application-driven approach requires capable, motivated, and self-sufficient students in order for the prototyping to succeed. Not every student was able to successfully finish by the end of the semester, particularly if the application-domain required students to learn multiple unfamiliar tools (e.g., robot operating system (ROS)) and design paradigms (e.g., 3D printing) that were not covered in the course. Therefore, in some cases, particularly for ambitious projects where students had to cover a lot of background, or projects with unplanned delays (e.g., delays in components being shipped, unexpected component failures, underperforming team member(s)), an extension was granted beyond the end of the semester, typically a month or less, for the students to finish their final prototype. In other cases, students who were not successful with the application-driven projects (during the proposal or mid-project review phases) were allowed to switch to survey/report-based projects.

## 5. FUTURE DIRECTIONS

There are many outstanding challenges in embedded systems education that must be addressed in the coming years:

**Data-driven and AI-based systems:** The growing importance of AI and machine learning in various embedded systems applications must be reflected in emerging embedded systems curricula. However, the demands and competencies required to understand AI and machine learning content, e.g., related to calculus, linear algebra, algorithm design, etc. are not equally (or adequately) represented in EE and CS curricula. Moreover, the principles at the intersection of embedded systems and AI are not well defined.

**Real engineering skills:** Many industry-relevant skills are hard to teach in courses, e.g., the ability to debug complex hardware and software components and their subsystems, understand poorly-written documentation, working in large teams and dynamics of managing large embedded design projects, coping with changing product specifications during the design process, etc. New approaches are needed to effectively cover such themes.

**Systems engineering integration:** Systems engineering is an engineering discipline that focuses on the challenge of managing development of complex technical systems. It originated in the aerospace and defense industry and has since been applied in other domains, e.g., transportation and energy grid. This discipline considers the complete system lifecycle from initial studies over development and manufacturing to operation and finally retirement. In doing so, this discipline considers many aspects that are ignored in embedded systems curricula today, such as understanding stakeholder needs, system operational and maintenance concepts, and overall lifecycle management [27]. The design of embedded systems curricula in the future can benefit from best practices of strong relevance and systems thinking that is the foundation of the systems engineering discipline.

**Ethics and diversity:** Ethical engineering is an important emerging challenge. Many examples exist of negative social impacts of technology, such as facial recognition software that discriminates against people of color [28]. As embedded systems become an integral part of our activities of daily life, playing a central role in how we work, learn, communicate, socialize, and participate in government, there is a need to train future embedded systems designers on the societal implications of technology, and to habituate them into thinking ethically as they develop algorithms and build embedded systems. Developers of new technologies need to try to identify potential harmful consequences early in the design process and take steps to eliminate or mitigate them [29]. This task is not easy. Embedded system designers will often have to negotiate among competing values, e.g., between efficiency and accessibility for a diverse user population, or between maximizing performance. There is no simple recipe for identifying and solving ethical problems. The field of embedded systems has historically also suffered from a lack of gender diversity. More efforts are needed to train future embedded system designers in the importance of ethics, diversity, equity, and inclusivity in technology.

## 6. CONCLUSION

In this paper, I discussed the evolution of embedded system education over the past two decades. I then outlined many outstanding pedagogical challenges in embedded systems education. Through more than a decade of experience with teaching embedded systems at Colorado State University, I presented a reflection on an application-centric approach to teaching embedded systems. This approach has shown promise to overcome some of the challenges facing embedded systems education. Lastly, looking into the future, there are several directions that embedded systems courses must consider as they evolve, and I discussed some themes of relevance for the designers of future embedded system curricula.